\documentclass[aps,pra,reprint,superscriptaddress,10pt, longbibliography]{revtex4-1}
\usepackage{amsmath}
\usepackage{hyperref}
\usepackage{graphicx}

\begin{document}

\title{Observation of an atom's electric quadrupole polarizability}

\author{Gerard Higgins}
\email[]{gerard.higgins@fysik.su.se}
\thanks{These two authors contributed equally.}
\affiliation{Department of Physics, Stockholm University, SE-106 91 Stockholm,
Sweden}
\author{Chi Zhang}
\email[]{chi.zhang@fysik.su.se}
\thanks{These two authors contributed equally.}
\affiliation{Department of Physics, Stockholm University, SE-106 91 Stockholm,
Sweden}
\author{Fabian Pokorny}
\affiliation{Department of Physics, Stockholm University, SE-106 91 Stockholm,
Sweden}
\author{Harry Parke}
\affiliation{Department of Physics, Stockholm University, SE-106 91 Stockholm,
Sweden}
\author{Erik Jansson}
\affiliation{Department of Physics, Stockholm University, SE-106 91 Stockholm,
Sweden}
\author{Shalina Salim}
\affiliation{Department of Physics, Stockholm University, SE-106 91 Stockholm,
Sweden}
\author{Markus Hennrich}
\email[]{markus.hennrich@fysik.su.se}
\affiliation{Department of Physics, Stockholm University, SE-106 91 Stockholm,
Sweden}


\date{\today}

\begin{abstract} 
The response of matter to fields underlies the physical sciences, from particle physics to astrophysics, and from chemistry to biophysics.
We observe an atom's response to an electric quadrupole field to second- and higher orders; this arises from the atom's electric quadrupole polarizability and hyperpolarizabilities.
We probe a single atomic ion which is excited to Rydberg states and confined in the electric fields of a Paul trap.
The quadrupolar trapping fields cause atomic energy level shifts and give rise to spectral sidebands.
The observed effects are described well by theory calculations.
\end{abstract}
\maketitle
Physics beyond the Standard Model is being probed by ever more precise spectroscopy \cite{Safronova2018}.
Also the continual exponential improvement of atomic clocks enables new technologies which impact society at large \cite{Ludlow2015}.
The quest for more precision requires ever more accurate determination of shifts due to external fields.
In this work we study second- and higher-order effects of electric quadrupole fields.

Second- and higher-order quadrupole responses contribute to inter-atomic \cite{Weber2017, Han2018} and inter-molecular interactions \cite{Gray1984}, and they have been observed as resonance shifts in bulk materials \cite{Cohen1957}.
Quadrupole effects are well-known in ion trapping systems, where strong electric quadrupole fields provide confinement.
First-order quadrupole shifts (due to the relatively weak static quadrupole field) affect trapped ion atomic clocks and precision spectroscopy experiments \cite{Itano2000}.
Second-order shifts (due to all of the trapping quadrupole fields) will become relevant in improved trapped ion precision experiments \cite{Ludlow2015}, particularly in molecular ion experiments, since molecules typically have larger quadrupole polarizabilities than atoms.

Various experiments are emerging which involve highly-excited Rydberg states in strong quadrupole trapping fields:
hybrid systems of neutral Rydberg atoms and trapped ions allow atom-ion interactions to be studied \cite{Ewald2019, Haze2019}, Rydberg molecules can be sensitively detected using an ion trap \cite{Deiss2020}, trapped Rydberg ions have recently shown potential for scalable quantum computing \cite{Zhang2020} and precision spectroscopy of Rydberg systems could be used to search for a fifth fundamental force \cite{Jones2020}.
The ion trap's electric fields cause extreme Stark effects in these systems \cite{Ewald2019, Haze2019, Mueller2008, Feldker2015, Higgins2017a, Higgins2019}; however, in this work we are concerned with the quadrupole effects that appear.

Rydberg states in ion traps can display giant first-order quadrupole effects \cite{Mueller2008, Higgins2017a}.
These effects are avoided by using $J=\frac{1}{2}$ Rydberg states, which do not have quadrupole moments.
However, quadrupole effects cannot be removed entirely; here we investigate second- and higher-order quadrupole effects displayed by a trapped ion in $J=\tfrac{1}{2}$ Rydberg states due to the states' quadrupole polarizabilities and hyperpolarizabilities.
We observe that the quadrupole trapping fields cause energy level shifts, and that the oscillating energy level shift due to the oscillating trapping field gives rise to spectral sidebands.
This oscillating energy opens new possibilities for Floquet engineering \cite{Oka2019}.
These effects will need to be considered in future experiments involving neutral Rydberg atoms, molecules or ions in electric quadrupole traps, as well as in future trapped ion clocks or precision spectroscopy experiments.

An external electric potential $\Phi$ perturbs a distribution of charges by
\begin{equation}
H' = \sum_i q_i \Phi = \sum_{ilm} q_i \phi_l^m r^l Y_l^m
\end{equation}
where $q_i$ are the charges, $\Phi$ is expanded in terms of spherical harmonics $Y_l^m$, and $\phi_l^m$ are scalars.
The effect of $H'$ on a state may be described perturbatively \cite{Gray1976}
\begin{equation}
E' = - \sum_{lm} Q_l^m \phi_l^m -\sum_{l_1l_2m_1m_2} \alpha_{l_1l_2}^{m_1m_2} \phi_{l_1}^{m_1} \phi_{l_2}^{m_2} + \mathcal{O}(\phi^3)
\end{equation}
where $Q_l^m$ is the state's $l^{\mathrm{th}}$ multipole moment, $\alpha_{l_1l_2}^{m_1m_2}$ is the multipole polarizability and the $\mathcal{O}$ represents higher-order terms.
Formulae describing $Q_l^m$, and $\alpha_{l_1l_2}^{m_1m_2}$ are given in the Supplemental Material \cite{appendix}. 

$Q_1^m$ describes the dipole moment, which gives rise to the Stark effect, and $\alpha_{11}^{m_1m_2}$ describes the dipole polarizability, which gives rise to the second-order (quadratic) Stark effect.
$Q_2^m$ describes the quadrupole moment, which gives rise to the first-order quadrupole shift, which is well-known in trapped ion precision spectroscopy.
In this work we present second-order quadrupole effects on an atom, which arise from the action of a quadrupole field on the atom's quadrupole polarizability $\alpha_{22}^{m_1m_2}$.

In our experiment the perturbing electric potential is the trapping potential of a linear Paul trap:
\begin{equation}
\Phi = (\phi_{\mathrm{rf}} \cos{\Omega t} + \phi_{\mathrm{rad}}) r^2 (Y_2^2 + Y_2^{-2}) + \phi_{\mathrm{ax}} r^2 Y_2^0
\end{equation}
where $\Omega$ is the frequency of the oscillating quadrupole field, $\phi_{\mathrm{rf}}$ describes its strength, the $\phi_{\mathrm{ax}}$ term provides axial confinement, the $\phi_{\mathrm{rad}}$ term causes the radial non-degeneracy, and the trap symmetry axis is collinear with the magnetic field axis.
These quantities are related to the ion's secular frequencies in the Supplemental Material \cite{appendix}.

In this potential a $J \leq \tfrac{1}{2}$ state, which has a quadrupole polarizability $\alpha_{22} \equiv \alpha_{22}^{00} = -\tfrac{1}{2}(\alpha_{22}^{1 -1} + \alpha_{00}^{-1 1}) = \tfrac{1}{2}(\alpha_{22}^{2 -2} + \alpha_{22}^{-2 2})$ and no quadrupole moment ($Q_2^m=0$), is shifted by
\begin{align}  \label{eq_energy_shift}
\begin{split}
E' &= \alpha_{22} \big( \phi_{\mathrm{rf}}^2 \cos{2 \Omega t} + 4 \phi_{\mathrm{rf}} \phi_{\mathrm{rad}} \cos{\Omega t} \\
& \qquad \quad + \phi_{\mathrm{rf}}^2 + 2 \phi_{\mathrm{rad}}^2 + \phi_{\mathrm{ax}}^2\big) + \mathcal{O}(\phi^3) \\
& \approx E'_2 \cos{2 \Omega t} + E'_1 \cos{\Omega t} + E'_0
\end{split}
\end{align}
where the $\mathcal{O}$ represents the higher-order response to quadrupole fields, and we introduce $E'_0$, $E'_1$ and $E'_2$ for clarity.
A more general equation describing general atomic states in terms of $Q_2^m$ and $\alpha_{22}^{m_1m_2}$ is given in the Supplemental Material \cite{appendix}.

The static energy shift $E'_0$ causes resonance frequency shifts, while the oscillating energy shifts $E'_1$ and $E'_2$ give rise to spectral sidebands.
We investigate both of these effects using spectroscopy of Rydberg $S_{1/2}$ and $P_{1/2}$ states of a single trapped $\mathrm{^{88}Sr^+}$ ion.
We excite a Zeeman sublevel of a Rydberg $S_{1/2}$ state by driving a two-UV-photon transition starting from metastable state $4D_{5/2}$ with intermediate state $6P_{3/2}$.
We probe a Zeeman sublevel of a Rydberg $P_{1/2}$ state using three fields; the two UV fields as well as a microwave field which couples Rydberg $S_{1/2}$ states to Rydberg $P_{1/2}$ states.
The Rydberg states decay overwhelmingly to the ground state $5S_{1/2}$, where detection is accomplished using ion fluorescence.
Further details about the experimental setup and the detection scheme can be found in \cite{Higgins2017a}.
We use a coherent spectroscopy technique to probe Rydberg $P_{1/2}$ states.
This technique offers resistance to double ionization losses \cite{Higgins2019_springer} and it is described in detail in the Supplemental Material \cite{appendix}.

Although the initial $4D_{5/2}$ state and intermediate $6P_{3/2}$ state have permanent quadrupole moments we safely neglect the effects of the quadrupole fields on these states, since the effects on the $J=\frac{1}{2}$ Rydberg states are around six orders of magnitude larger.

\begin{figure}[ht!]
\centering
\includegraphics[width=\columnwidth]{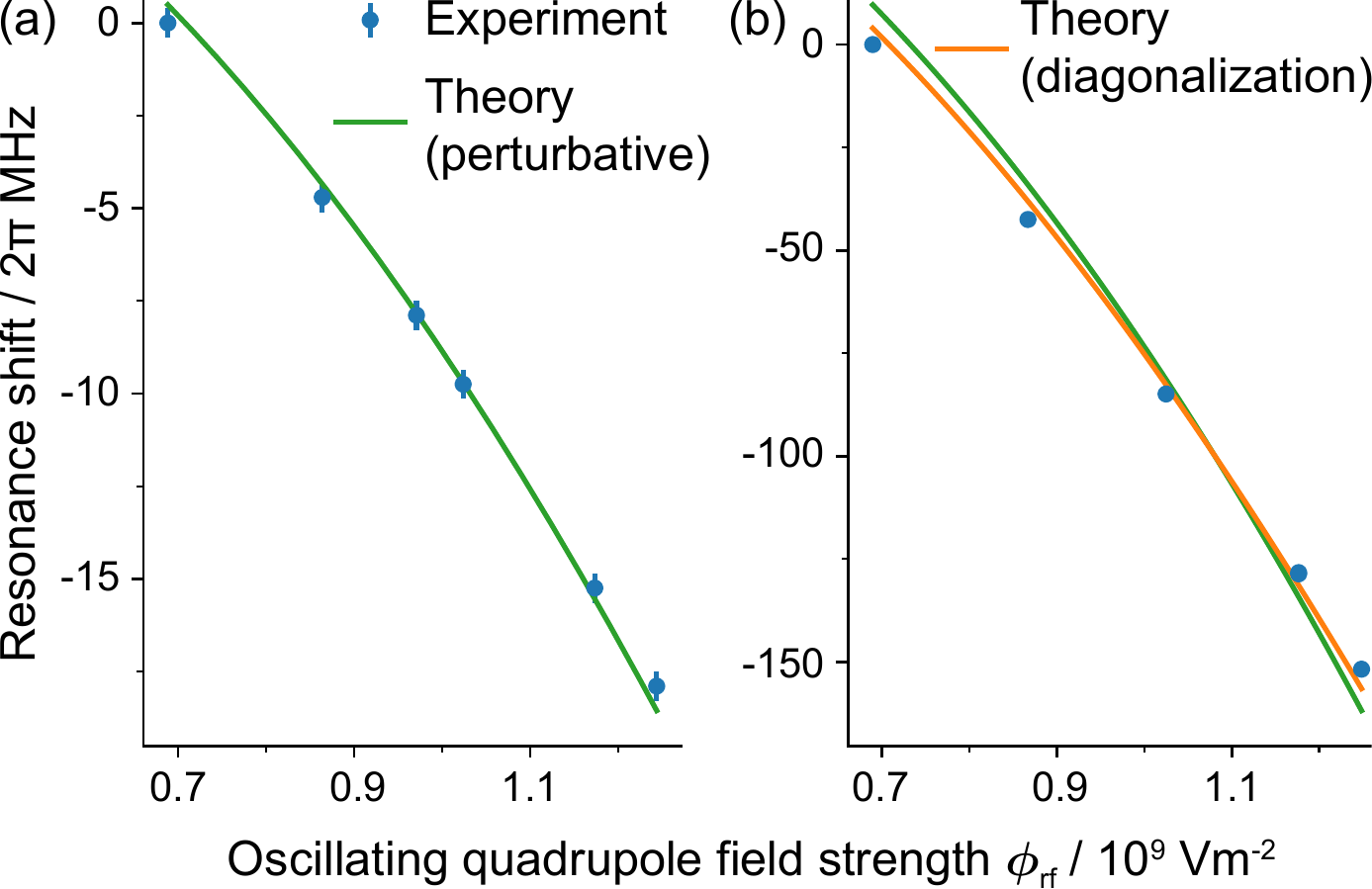}
\caption{Rydberg state energies depend on the strength of the oscillating quadrupole field $\phi_{\mathrm{rf}}$.
(a)~The response of $56S_{1/2}$ to $\phi_{\mathrm{rf}}$ is described by the second-order perturbative calculation.
(b)~The response of $56P_{1/2}$ to $\phi_{\mathrm{rf}}$ is described better by the full diagonalization calculation than by the second-order perturbative calculation.
Uncertainties in resonance frequencies are represented by error bars (68\% confidence intervals).
The error bars in (b) are smaller than the markers.
}
\label{fig1}
\end{figure}

The static energy level shift due to the quadrupole fields ($E'_0$ in Eq.~\ref{eq_energy_shift}) means that the energy required to excite Rydberg states depends on the field strengths.
We investigate this by measuring the $4D_{5/2} \leftrightarrow 56S_{1/2}$ and $4D_{5/2} \leftrightarrow 56P_{1/2}$ resonance frequencies as the amplitude of the oscillating field $\phi_{\mathrm{rf}}$ is varied; the results are shown in Fig.~\ref{fig1}.
The energy required for excitation of $56S_{1/2}$ depends quadratically on $\phi_{\mathrm{rf}}$ [Fig.~\ref{fig1}(a)], according to the second-order perturbation description in Eq.~\ref{eq_energy_shift}.
The theory curve uses the theory value of the $56S_{1/2}$ quadrupole polarizability $\alpha_{22}$, the calculation is described in detail in the Supplemental Material \cite{appendix}.

Rydberg $P_{1/2}$ states are generally more sensitive to quadrupole fields than Rydberg $S_{1/2}$ states, since the fields couple $P_{1/2}$ states to the energetically nearby $P_{3/2}$ states (as well as more distant $F_{5/2}$ states), while the fields only couple $S_{1/2}$ states to the more distant $D$ states.
The higher sensitivity of $56P_{1/2}$ relative to $56S_{1/2}$ is seen by comparing Fig.~\ref{fig1}(b) and (a).
The strong response of $56P_{1/2}$ to $\phi_{\mathrm{rf}}$ means that higher-order terms cannot be neglected.
We capture these higher-order terms by full diagonalization of the Hamiltonian (details in \cite{appendix}) which describes the experimental data better than the second-order perturbation calculation.

\begin{figure}[ht!]
\centering
\includegraphics[width=\columnwidth]{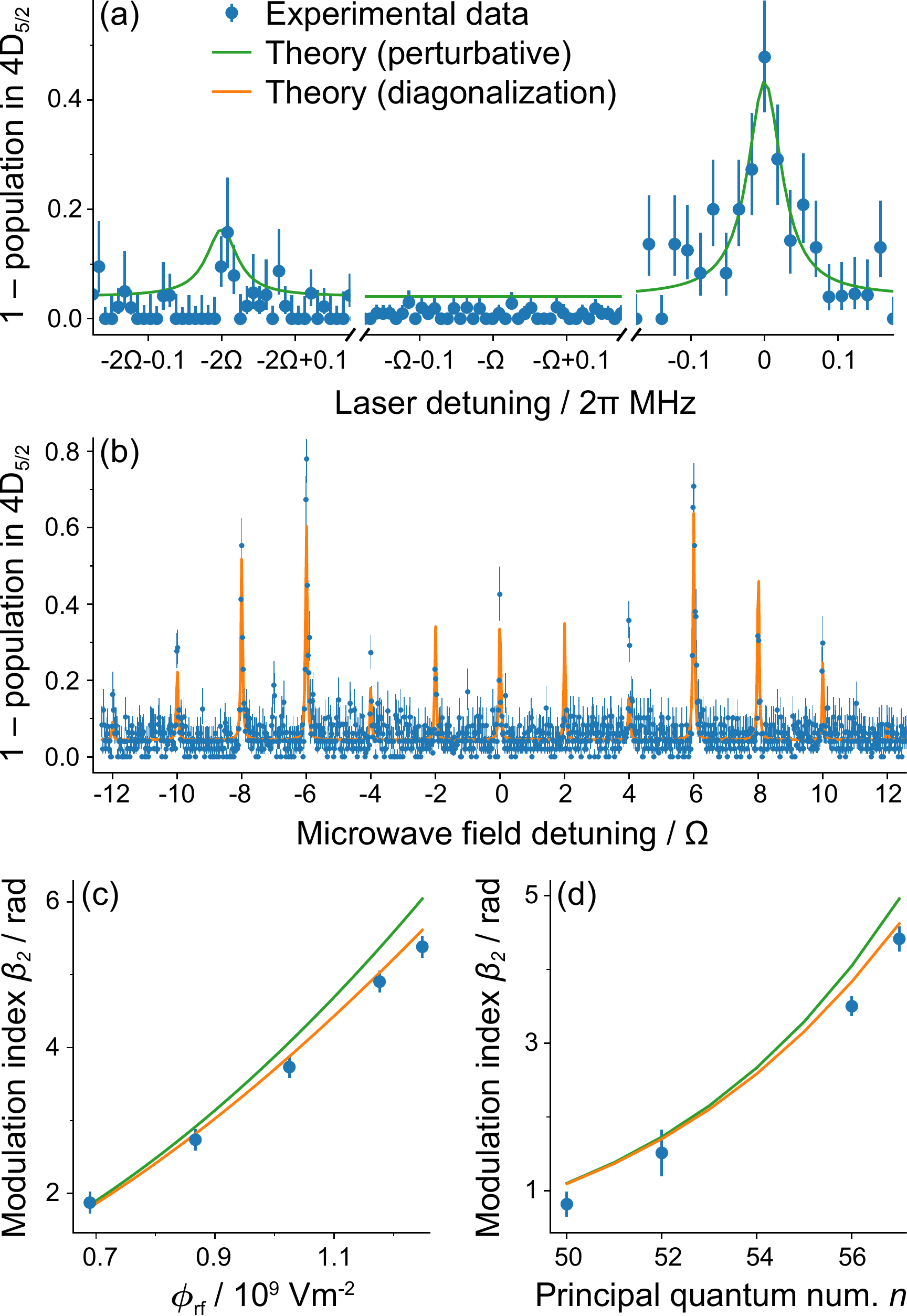}
\caption{The oscillating quadrupole shift introduces sidebands to Rydberg-excitation spectra.
(a)~The $56S_{1/2}$ excitation spectrum displays weak 2$\Omega$ sidebands.
(b)~The $57P_{1/2}$ excitation spectrum displays a forest of sidebands.
(c)~The modulation index $\beta_2$ (which describes the sideband amplitudes) increases with the amplitude of the oscillating quadrupole field $\phi_{\mathrm{rf}}$; state $56P_{1/2}$ was used.
(d)~$\beta_2$ increases with the principal quantum number $n$ of Rydberg $P_{1/2}$ states.
In (c) and (d) the measured $\beta_2$ is described better by the full diagonalization calculation than by the second-order perturbative calculation.
Throughout this work $\Omega=2\pi\times 18.1\,\mathrm{MHz}$.
Error bars represent 68\% confidence intervals; in (a) and (b) error bars represent quantum projection noise, in (c) and (d) error bars represent uncertainties in extracting $\beta_2$ from spectra.
}
\label{fig2}
\end{figure}

The oscillating quadrupole field $\phi_{\mathrm{rf}}$ causes an oscillating quadrupole shift (Eq.~\ref{eq_energy_shift}), which leads to sidebands in Rydberg-excitation spectra, as shown in Fig.~\ref{fig2}(a) and (b).
Two sets of sidebands appear: the modulation with frequency $\Omega$ causes sidebands at multiples of $\Omega$ described by modulation index $\beta_1 = \frac{E'_1}{\hbar \Omega}$ while the modulation with frequency $2\Omega$ causes sidebands at multiples of $2\Omega$ described by modulation index $\beta_2 = \frac{E'_2}{2\hbar \Omega}$.
In a linear Paul trap the oscillating field strength is usually much larger than the strengths of the static fields $\phi_{\mathrm{rf}} \gg \phi_{\mathrm{ax}}, \phi_{\mathrm{rad}}$ and $E'_0 \approx E'_2 \gg E'_1$ and $\beta_2 \gg \beta_1$.
This means the sidebands at multiples of $2\Omega$ are much larger than the sidebands at multiples of $\Omega$.
We operate our trap in this regime.
Spectral sidebands emerge when the excitation timescale is much longer than the modulation period, as described in the Supplemental Material \cite{appendix}.

The $4D_{5/2} \leftrightarrow 56S_{1/2}$ spectrum in Fig.~\ref{fig2}(a) displays small sidebands due to the $2\Omega$ modulation while no sidebands due to the weaker $\Omega$ modulation are visible.
Good agreement is observed between the experimental data and the theory curve: the theory peak at $-2\Omega$ is constrained by the calculated $56S_{1/2}$ quadrupole polarizability, the measured quadrupole field strengths, and the fit of the carrier resonance.

Before we discuss $nP_{1/2}$ spectra we refine a point made earlier: Due to the coherent spectroscopy technique we use to probe $nP_{1/2}$ states, the $nP_{1/2}$ excitation spectra are described by a modulation index which depends on the difference between the $nS_{1/2}$ and $nP_{1/2}$ modulations $\beta_2 = \frac{E'_{2,P} - E'_{2,S}}{2\hbar \Omega}$.
This is corroborated by simulations in \cite{appendix}.

$57P_{1/2}$ is much more sensitive to quadrupole fields than $56S_{1/2}$; using the same trap settings as in Fig.~\ref{fig2}(a) the $4D_{5/2} \leftrightarrow 57P_{1/2}$ spectrum appears as a forest of resonance peaks, shown in Fig.~\ref{fig2}(b).
The relative strengths of the resonances are well-described by Bessel functions of the first kind $\left[J_m\left(\beta_2\right)\right]^2$ at the $m^{\mathrm{th}}$ multiple of $2\Omega$, with $\beta_2=4.62\,\mathrm{rad}$.
This value of $\beta_2$ was found by full diagonalization of the Hamiltonian (details in \cite{appendix}).
The experimental data is described well by the theory curve.
The only free parameters in the theory curve are the centre frequency, the effective excitation Rabi frequency, the resonance linewidth, the background level and the signal's saturation point.

The $\Omega$ modulation gives rise to weaker sidebands at multiples of $\Omega$.
The interplay between the $\Omega$ modulation and $2\Omega$ modulation causes a slight asymmetry in the spectrum, which is explained in the Supplemental Material \cite{appendix}.
A similar interplay of $\Omega$ and $2\Omega$ modulations due to Doppler and Stark effects was observed in \cite{Feldker2015}.

$56P_{1/2}$ excitation spectra were measured as the amplitude of the oscillating quadrupole field $\phi_{\mathrm{rf}}$ was varied.
The modulation index $\beta_2$ was extracted from the spectra and its dependence on $\phi_{\mathrm{rf}}$ is shown in Fig.~\ref{fig2}(c).
The full diagonalization calculation describes the experimental data better than the second-order perturbation calculation, as was the case for the data in Fig.~\ref{fig1}(b).

Excitation spectra of Rydberg $P_{1/2}$ states with principal quantum numbers $n$ between 50 and 57 were measured; the quadrupole field strengths were kept fixed.
$\beta_2$ values were extracted and the dependence of $\beta_2$ on $n$ is shown in Fig.~\ref{fig2}(d).
The response of Rydberg $P_{1/2}$ states to quadrupole fields is captured better by the full diagonalization calculation than by the second-order perturbative approach, just as for the data in Fig.~\ref{fig1}(b) and Fig.~\ref{fig2}(c).
The perturbation theory curve scales with the effective principal quantum number $n^*$ as ${n^*}^{11}$; this is the expected scaling of the quadrupole polarizability (this comes from second-order perturbation theory: quadrupole couplings grow with ${n^*}^{4}$ and energy splittings decay as ${n^*}^{-3}$).

The oscillating energy shift (Eq.~\ref{eq_energy_shift}) is challenging to quantitatively investigate when the modulation index $\beta$ is small, as is the case for Rydberg $S_{1/2}$ states.
Using spectroscopy the height of the first sideband is smaller than the height of the carrier by $\left[ \frac{J_1(\beta)}{J_0(\beta)} \right]^2 \approx \frac{\beta^2}{4}$; this is the case in Fig.~\ref{fig2}(a).
Coherent spectroscopy techniques can offer a more favourable approach when $\beta$ is small; with coherent spectroscopy the effect of the first sideband transition is smaller than the effect of the carrier by $\frac{J_1(\beta)}{J_0(\beta)} \approx \frac{\beta}{2}$.

\begin{figure}[ht!]
\centering
\includegraphics[width=\columnwidth]{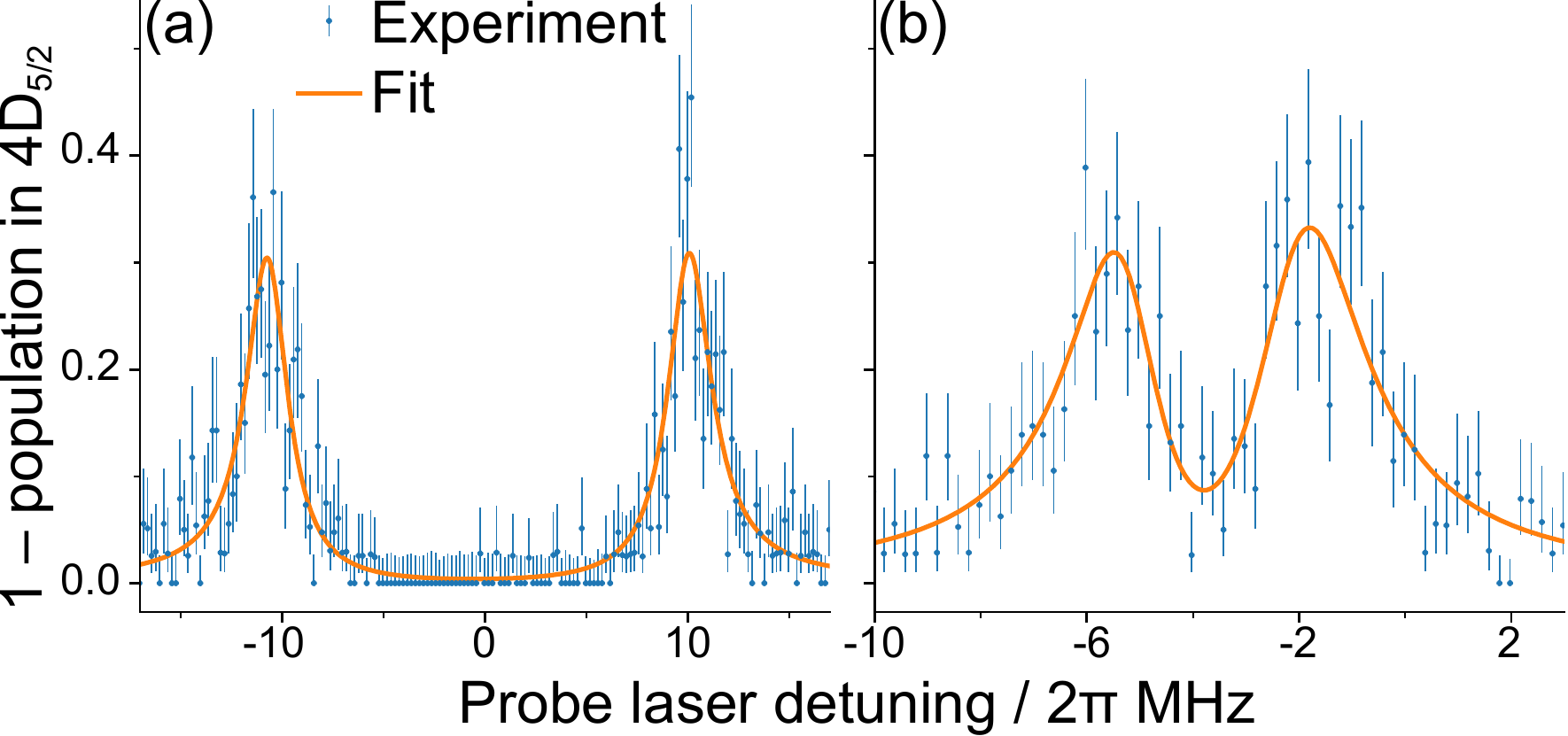}
\caption{The oscillating quadrupole shift is probed using the Autler-Townes effect.
The oscillating quadrupole shift introduces sidebands $\pm 2 \Omega$ from the $6P_{3/2} \leftrightarrow 46S_{1/2}$ resonance.
The sideband coupling strength is compared with the carrier coupling strength by comparing the Autler-Townes splittings that they each induce on the weakly-probed $4D_{5/2} \leftrightarrow 6P_{3/2}$ transition.
In (a) the coupling laser is resonant to the $6P_{3/2} \leftrightarrow 46S_{1/2}$ carrier transition, in (b) the coupling laser is resonant to a sideband transition.
Error bars represent quantum projection noise (68\% confidence intervals).
}
\label{fig3}
\end{figure}

We use the Autler-Townes effect to probe weak modulations, as shown in Fig.~\ref{fig3}.
We excite Rydberg $S_{1/2} \equiv |2\rangle$ states with a two-photon transition from $4D_{5/2} \equiv |0\rangle$ via the intermediate $6P_{3/2} \equiv |1\rangle$ state; with this transition the Autler-Townes effect can be observed \cite{Higgins2017b}.
If the second laser field resonantly couples $|1\rangle$ and $|2\rangle$ with strength $J_0(\beta) \Omega_c$, then the $|0\rangle \leftrightarrow |1\rangle$ excitation spectrum displays two resonance peaks split by $J_0(\beta) \Omega_c$ [see Fig.~\ref{fig3}(a)].
If the second laser field is detuned such that it couples $|1\rangle$ and $|2\rangle$ via the first sideband transition, this coupling then has strength $J_1(\beta) \Omega_c$, and the peaks in the $|0\rangle \leftrightarrow |1\rangle$ spectrum are split by $J_1(\beta) \Omega_c$ [see Fig.~\ref{fig3}(b)].
The ratio of the two splittings is $\frac{J_1(\beta)}{J_0(\beta)} \approx \frac{\beta}{2}$ (the approximation is valid for $\beta \ll 1$).
From such measurements we determine the $46S_{1/2}$ quadrupole polarizability $\alpha_{22} = (1.08 \pm 0.05) \times 10^{-45}\,\mathrm{J m^4 V^{-2}}$, which is similar to the calculated value $\alpha_{22} = 1.21 \times 10^{-45}\,\mathrm{J m^4 V^{-2}}$.

If the strengths of the trapping quadrupole fields drift in time this will cause quadrupole shifts ($E'_0$) and resonance frequencies to drift.
This will impair precision spectroscopy experiments as well as experiments using sensitive Rydberg states in strong quadrupole fields.
We combat drifts by actively stabilising the amplitude of the oscillating electric field in our system \cite{Hempel2014, Johnson2016}.

Systems of trapped Rydberg ions have recently shown great potential as a scalable platform for quantum computation and simulation \cite{Zhang2020}.
However, unwanted coupling to quadrupole-field-induced sidebands may reduce the fidelity of a Rydberg ion quantum gate.
This unwanted coupling may be reduced by appropriate choice of $\phi_{\mathrm{rf}}$ such that $J_{\pm 1}(\beta_2)=0$ and the first-order sidebands at $\pm 2 \Omega$ vanish, then the nearest significant sidebands would be detuned by $\pm 4 \Omega$.
Alternatively the coupling to sidebands may be diminished by increasing $\Omega$ \cite{Simeonov2019}.
One could remove the sidebands entirely by modulating the Rydberg-excitation laser fields and microwave field such that these fields follow the oscillating Rydberg energy levels, or else one could confine ions using a rotating Paul trap \cite{Hasegawa2005}, a Penning trap \cite{Brown1986} or a digital ion trap \cite{Deb2015, Haze2019} instead of an oscillating Paul trap.

In this work we investigate higher-order effects of electric quadrupole fields on a single atom.
Effects on Rydberg $S_{1/2}$ states are described well in terms of the electric quadrupole polarizability.
Rydberg $P_{1/2}$ states are more sensitive to quadrupole fields, and full diagonalization calculations describe the response of $P_{1/2}$ states better than the second-order perturbation calculations.
The resonance shifts and spectral sidebands we observe will need to be considered in future experiments involving highly-sensitive Rydberg atoms, molecules or ions in ion traps.
Additionally the resonance shifts will be important in future trapped ion precision experiments and clocks \cite{Ludlow2015}, particularly in experiments involving molecular ions.

The huge \textit{dipole} polarizabilities of trapped Rydberg ions offer a range of applications \cite{Li2012, Li2013, Nath2015, Vogel2019, Higgins2019, Gambetta2019}.
We hope that this work will stimulate new research avenues which take advantage of the extreme \textit{quadrupole} polarizabilities of trapped Rydberg ions; for instance, the sensitivity of Rydberg ions to oscillating quadrupole fields may allow Floquet engineering \cite{Oka2019}.
We also hope to stimulate investigation of the quadrupole map -- the quadrupole field counterpart to the Stark map \cite{Zimmerman1979}.

A final remark:
We determine the $\mathrm{^{88}Sr^+}$ $56S_{1/2}$ quadrupole polarizability to be $\alpha_{22}\approx 10^{-44}\,\mathrm{J m^4 V^{-2}}\approx10^{70}\,\mathrm{a.u.}$ -- it is quite striking that SI units are better suited for expressing this atomic state property than atomic units!
This exemplifies the macroscopicity of Rydberg atoms.

\subsection*{Acknowledgements}
We thank Cornelius Hempel for information about trap stabilization, Shinsuke Haze and Markus Dei{\ss} for information about hybrid atom-ion systems and digital ion traps, and all members of the ERyQSenS consortium for discussions.
This work was supported by the European Research Council under the European Unions Seventh Framework Programme/ERC Grant Agreement No. 279508, the Swedish Research Council (Trapped Rydberg Ion Quantum Simulator), the QuantERA ERA-NET Cofund in Quantum Technologies (ERyQSenS), and the Knut \& Alice Wallenberg Foundation (Photonic Quantum Information and WACQT).

\subsection*{Author contributions}
CZ first measured and interpreted the resonance shift, GH and CZ carried out the calculations, CZ developed the spectroscopy technique for Rydberg $P$ states, GH, FP, CZ, HP, EJ and SS carried out the measurements, GH wrote the manuscript, MH supervised the project, all authors contributed to the discussion of the results and to the manuscript.


%

\newcommand{\snum}{S}

\setcounter{equation}{0}
\setcounter{figure}{0}
\renewcommand{\theequation}{\snum \arabic{equation}}
\renewcommand{\thefigure}{\snum \arabic{figure}}

\onecolumngrid

\section*{Supplemental material}



\section{Calculation of the multipole moment, multipole polarizability and first multipole hyperpolarizability using perturbation theory} \label{sec_a}
For an atom with a single valence electron, and following a perturbative approach \cite{Gray1976}:
\begin{align}
Q_l^m &= e \langle \psi | r^l Y_l^m | \psi \rangle \\
\alpha_{l_1l_2}^{m_1m_2} &= e^2 \sum_{\psi'} \frac{ \langle \psi | r^{l_1} Y_{l_1}^{m_1} | \psi' \rangle \langle \psi' | r^{l_2} Y_{l_2}^{m_2} | \psi \rangle }{E_{\psi'} - E_{\psi}} \label{eq_second_order_pert} \\
\beta_{l_1l_2l_3}^{m_1m_2m_3} &= e^3 \sum_{\psi'} \sum_{\psi''} \frac{ \langle \psi | r^{l_1} Y_{l_1}^{m_1} | \psi' \rangle \langle \psi' | r^{l_2} Y_{l_2}^{m_2} | \psi'' \rangle \langle \psi'' | r^{l_3} Y_{l_3}^{m_3} | \psi \rangle }{\left(E_{\psi'} - E_{\psi}\right) \left(E_{\psi''} - E_{\psi}\right)} - e^3 \langle \psi | r^{l_1} Y_{l_1}^{m_1} | \psi \rangle \sum_{\psi'} \frac{ \langle \psi | r^{l_2} Y_{l_2}^{m_2} | \psi' \rangle \langle \psi' | r^{l_3} Y_{l_3}^{m_3} | \psi \rangle }{(E_{\psi'} - E_{\psi})^2}
\end{align}
where $\beta_{l_1l_2l_3}^{m_1m_2m_3}$ is the first multipole hyperpolarizability.

An electronic wavefunction in the spin-orbit coupling regime is described by quantum numbers $\{n,L,S,J,m_J\}$.
The wavefunctions of atoms with a single valence electron can be separated into radial and angular parts.
\begin{align}
|\psi\rangle &= |nLSJm_J\rangle \\
&= |nLSJ\rangle |LSJm_J\rangle
\end{align}
The separability of the wavefunction is used to calculate the matrix elements of $e r^l Y_l^m$
\begin{equation}
\langle n_1 L_1 S_1 J_1 m_{J1} | e r^l Y_l^m | n_2 L_2 S_2 J_2 m_{J2} \rangle = e \langle n_1 L_1 S_1 J_1 | r^l | n_2 L_2 S_2 J_2 \rangle \langle L_1 S_1 J_1 m_{J1} | Y_l^m | L_2 S_2 J_2 m_{J2} \rangle
\end{equation}
The radial matrix elements are found numerically.
The angular matrix elements are found by expanding
\begin{align}
|LSJm_J\rangle &= \sum_{m_S} \sum_{m_L} \langle LSm_Lm_S|LSJm_J\rangle \:\: |LSm_Lm_S\rangle \:\: \delta_{m_L+m_S,m_J} \\
&= \sum_{m_S} \langle LS(m_J-m_S)m_S|LSJm_J\rangle \:\: |LS(m_J-m_S)m_S\rangle \\
&= \sum_{m_S} \langle LS(m_J-m_S)m_S|LSJm_J\rangle \:\: |Y_L^{m_J-m_S}\rangle \:\: |Sm_S\rangle
\end{align}
where $\langle LSm_Lm_S|LSJm_J\rangle$ is a Clebsch-Gordan coefficient
\begin{equation}
\langle LSm_Lm_S|LSJm_J\rangle = (-1)^{L-S+m_J} \sqrt{2J+1} \begin{pmatrix}
L & S & J \\
m_L & m_S & -m_J 
\end{pmatrix}
\end{equation}

The operator $e r^l Y_l^m$ does not affect the spin state, thus the non-zero matrix elements have $S_1 = S_2 = S$.
Using $\langle S m_{S1} | S m_{S2} \rangle = \delta_{m_{S1},m_{S2}}$ we can now write the angular matrix element
\begin{align}
\begin{split}
\langle L_1 S J_1 m_{J1} | Y_l^m | L_2 S J_2 m_{J2} \rangle = \sum_{m_S} &\langle L_1 S (m_{J1}-m_S) m_S | L_1 S J_1 m_{J1} \rangle \\
\times &\langle L_2 S (m_{J2}-m_S) m_S | L_2 S J_2 m_{J2} \rangle \\
\times &\langle Y_{L_1}^{m_{J1}-m_S} | Y_l^m | Y_{L_2}^{m_{J2}-m_S} \rangle
\end{split}
\end{align}
where $\langle Y_{L_1}^{m_{J1}-m_S} | Y_l^m | Y_{L_2}^{m_{J2}-m_S} \rangle$ is an integral of three spherical harmonics
\begin{align}
\langle Y_{L_1}^{m_1} | Y_l^m | Y_{L_2}^{m_2} \rangle &= \int {Y_{L_1}^{m_1}}^* Y_l^m Y_{L_2}^{m_2} d\Omega \\
&= (-1)^{m_1} \sqrt{\frac{(2L_1+1)(2l+1)(2L_2+1)}{4\pi}} \begin{pmatrix}
L_1 & l & L_2 \\
0 & 0 & 0 
\end{pmatrix}
\begin{pmatrix}
L_1 & l & L_2 \\
-m_1 & m & m_2 
\end{pmatrix} \label{eq_integral_three_spherical_harmonics}
\end{align}
For the right-most Wigner-3j symbol in Eq.~\ref{eq_integral_three_spherical_harmonics} to be non-zero, $-m_1 + m + m_2 = 0$.
As a result, the only non-zero elements of $Q_l^m$, $\alpha_{l_1l_2}^{m_1m_2}$ and $\beta_{l_1l_2l_3}^{m_1m_2m_3}$ have $m=0$, $m_1+m_2=0$ and $m_1+m_2+m_3=0$ respectively.

We calculated properties of $\mathrm{^{88}Sr^+}$ Rydberg states by adapting the Alkali Rydberg Calculator \cite{Sibalic2017}.

\section{Formulae relating the field gradients to the secular frequencies}
The trapping potential of a linear Paul trap is given by
\begin{align}
\Phi &= (\phi_{\mathrm{rf}} \cos{\Omega t} + \phi_{\mathrm{rad}}) r^2 (Y_2^2 + Y_2^{-2}) + \phi_{\mathrm{ax}} r^2 Y_2^0 \\
&= \frac{1}{2} \sqrt{\frac{15}{2\pi}} (\phi_{\mathrm{rf}} \cos{\Omega t} + \phi_{\mathrm{rad}}) (x^2 - y^2) - \frac{1}{4} \sqrt{\frac{5}{\pi}} \phi_{\mathrm{ax}} \left( x^2 + y^2 - 2 z^2 \right)
\end{align}
where $\Omega$ is the frequency of the oscillating field, $\phi_{\mathrm{rf}}$ describes its gradient, the $\phi_{\mathrm{ax}}$ term provides axial confinement and the $\phi_{\mathrm{rad}}$ term causes the radial non-degeneracy.

The secular frequencies of a trapped ion are related to the field gradients and the trap drive frequency by
\begin{align}
\omega_x^2 &= \frac{15 e^2 \phi_{\mathrm{rf}}^2}{4 \pi M^2 \Omega^2} - \sqrt{\frac{5}{\pi}} \frac{1}{2} \frac{e \phi_{\mathrm{ax}}}{M} + \sqrt{\frac{15}{2\pi}} \frac{e \phi_{\mathrm{rad}}}{M} \\
\omega_y^2 &= \frac{15 e^2 \phi_{\mathrm{rf}}^2}{4 \pi M^2 \Omega^2} - \sqrt{\frac{5}{\pi}} \frac{1}{2} \frac{e \phi_{\mathrm{ax}}}{M} - \sqrt{\frac{15}{2\pi}} \frac{e \phi_{\mathrm{rad}}}{M} \\
\omega_z^2 &= \sqrt{\frac{5}{\pi}} \frac{e \phi_{\mathrm{ax}}}{M}
\end{align}
where $M$ is the ion mass and $e$ is the elementary charge.

\begin{align}
\phi_{\mathrm{rf}} &= \sqrt{\frac{2\pi}{15}} \frac{1}{e} M \Omega \sqrt{ \omega_x^2 + \omega_y^2 + \omega_z^2 } \\
\phi_{\mathrm{ax}} &= \sqrt{\frac{\pi}{5}} \frac{1}{e} M \omega_z^2 \\
\phi_{\mathrm{rad}} &= \sqrt{\frac{\pi}{30}} \frac{1}{e} M (\omega_x^2 - \omega_y^2)
\end{align}

\section{Considering different orientations of the trap symmetry axis}
Eq.~3 in the main text describes how the quadrupole fields are decomposed when the trap symmetry axis is collinear with the quantization axis (defined by the magnetic field).
When these axes are not collinear one may rotate between the two coordinate systems defined by the magnetic field and by the trap axis using the Wigner D-matrix, as done in \cite{Itano2000}.

\section{Considering quadrupole polarizability tensors}
The $J=\frac{1}{2}$ states considered in this work have quadrupole polarizabilities which satisfy
\begin{equation}
\alpha_{22} \equiv \alpha_{22}^{00} = \tfrac{1}{2}(\alpha_{22}^{1-1}+\alpha_{22}^{-11}) = \tfrac{1}{2}(\alpha_{22}^{2-2} + \alpha_{22}^{-22})
\end{equation}
this simplified Eq.~4 considerably.

More generally (for the case where the quantization axis is aligned with the trap symmetry axis), and using the knowledge from section~\ref{sec_a} that the only non-zero elements of $Q_l^m$, $\alpha_{l_1l_2}^{m_1m_2}$ and $\beta_{l_1l_2l_3}^{m_1m_2m_3}$ have $m=0$, $m_1+m_2=0$ and $m_1+m_2+m_3=0$ respectively.
To third-order:
\begin{align}
\begin{split}
E' &= Q_2^0 \phi_{\mathrm{ax}} + \alpha_{22}^{00} (\phi_{\mathrm{ax}})^2 + (\alpha_{22}^{2-2}+\alpha_{22}^{-22}) (\phi_{\mathrm{rf}} \cos{\Omega t} + \phi_{\mathrm{rad}})^2 \\
& \:\:\:\:\: + \beta_{222}^{000} (\phi_{\mathrm{ax}})^3 + (\beta_{222}^{2-20}+\beta_{222}^{-220}+\beta_{222}^{20-2}+\beta_{222}^{-202}+\beta_{222}^{02-2}+\beta_{222}^{0-22}) \phi_{\mathrm{ax}} (\phi_{\mathrm{rf}} \cos{\Omega t} + \phi_{\mathrm{rad}})^2 \\
&=Q_2^0 \phi_{\mathrm{ax}} + \alpha_{22}^{00} (\phi_{\mathrm{ax}})^2 + \beta_{222}^{000} (\phi_{\mathrm{ax}})^3  \\
& \:\:\:\:\: + \left[ \alpha_{22}^{2-2}+\alpha_{22}^{-22} + (\beta_{222}^{2-20}+\beta_{222}^{-220}+\beta_{222}^{20-2}+\beta_{222}^{-202}+\beta_{222}^{02-2}+\beta_{222}^{0-22}) \phi_{\mathrm{ax}} \right] (\phi_{\mathrm{rf}} \cos{\Omega t} + \phi_{\mathrm{rad}})^2
\end{split} \label{eq_e_prime_supp}
\end{align}
The last part of the final term describes the time-dependence
\begin{align}
(\phi_{\mathrm{rf}} \cos{\Omega t} + \phi_{\mathrm{rad}})^2 = \tfrac{1}{2} \phi_{\mathrm{rf}}^2 \cos{2 \Omega t} + 2 \phi_{\mathrm{rf}} \phi_{\mathrm{rad}} \cos{\Omega t} + \phi_{\mathrm{rad}}^2 + \tfrac{1}{2} \phi_{\mathrm{rf}}^2
\end{align}
Thus Eq.~\ref{eq_e_prime_supp} can be simplified to
\begin{equation}
E' = E'_0 + E'_1 \cos{\Omega t} + E'_2 \cos{2 \Omega t}
\end{equation}
The $\Omega$ modulation and $2\Omega$ modulation have modulation indices
\begin{align}
\beta_1 &= \frac{E'_1}{\hbar \Omega} \\
&= \left[ \alpha_{22}^{2-2}+\alpha_{22}^{-22} + (\beta_{222}^{2-20}+\beta_{222}^{-220}+\beta_{222}^{20-2}+\beta_{222}^{-202}+\beta_{222}^{02-2}+\beta_{222}^{0-22}) \phi_{\mathrm{ax}} \right] \frac{2\phi_{\mathrm{rf}}\phi_{\mathrm{rad}}}{\hbar \Omega} \\
\beta_2 &= \frac{E'_2}{2 \hbar \Omega} \\
&= \left[ \alpha_{22}^{2-2}+\alpha_{22}^{-22} + (\beta_{222}^{2-20}+\beta_{222}^{-220}+\beta_{222}^{20-2}+\beta_{222}^{-202}+\beta_{222}^{02-2}+\beta_{222}^{0-22}) \phi_{\mathrm{ax}} \right] \frac{\phi_{\mathrm{rf}}^2}{4 \hbar \Omega}
\end{align}
whose ratio is
\begin{equation}
\frac{\beta_1}{\beta_2} = \frac{8\phi_{\mathrm{rad}}}{\phi_{\mathrm{rf}}}
\end{equation}
Usually $\phi_{\mathrm{rad}} \ll \phi_{\mathrm{rf}}$ and $\beta_1 \ll \beta_2$.

\section{Coherent spectroscopy technique} \label{sec_coherent_spectroscopy}
The $\mathrm{^{88}Sr^+}$ ion can loose its Rydberg electron in a process of double ionization $\mathrm{^{88}Sr^+} \rightarrow \mathrm{^{88}Sr^{2+}}+e^-$ \cite{Higgins2019_springer}.
The unbound electron quickly leaves the system, and a new $\mathrm{^{88}Sr^+}$ ion must be loaded for experiments to continue.
This slows down Rydberg ion experiments, and as a result, errors due to slow drifts of the system become more appreciable.
Our experience shows that the double ionization rate depends on the time that the ion spends in the Rydberg state.
The ionization rates that we observe are consistent with ionization caused by blackbody radiation in the room temperature trap while the ion is in the Rydberg state \cite{Zhang2020}.

By using the following technique we measure Rydberg-excitation spectra while being more resistant to double ionization loss.
Instead of exciting Rydberg states, we excite laser-dressed states which have a small Rydberg component and a significant component of state $6P_{3/2}$.
Because less population is excited to Rydberg states we reduce the likelihood of loosing ions by double ionization.
More importantly, because $6P_{3/2}$ has a short lifetime ($\approx 35\,\mathrm{ns}$), the lifetime of the laser-dressed state can be made much shorter than Rydberg state lifetimes ($\sim 10\,\mathrm{\mu s}$ for $n\sim50$).
This reduces the time the ion spends in the Rydberg state and thus the likelihood of double ionization loss.

\begin{figure}[ht!]
\centering
\includegraphics[width=0.5\columnwidth]{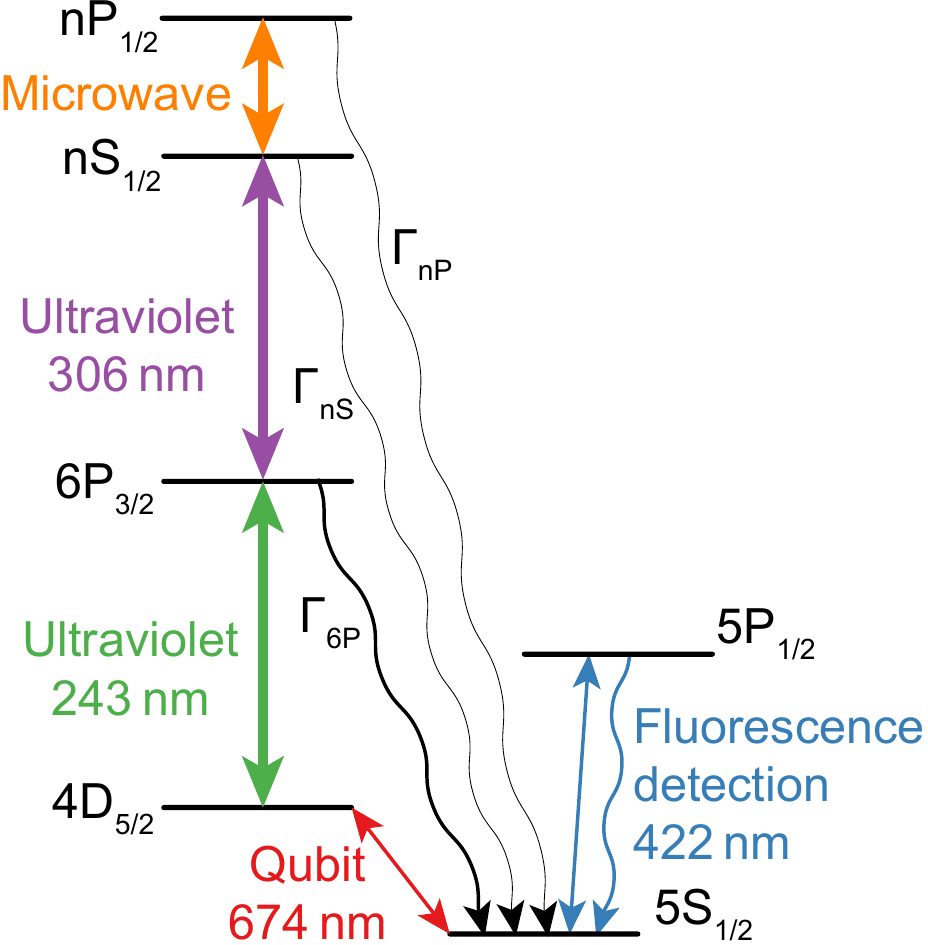}
\caption{Relevant level scheme of $\mathrm{^{88}Sr^+}$.
}
\label{fig_supp_level_scheme}
\end{figure}

The level scheme is shown in Fig.~\ref{fig_supp_level_scheme}.

\begin{figure}[ht!]
\centering
\includegraphics[width=0.8\columnwidth]{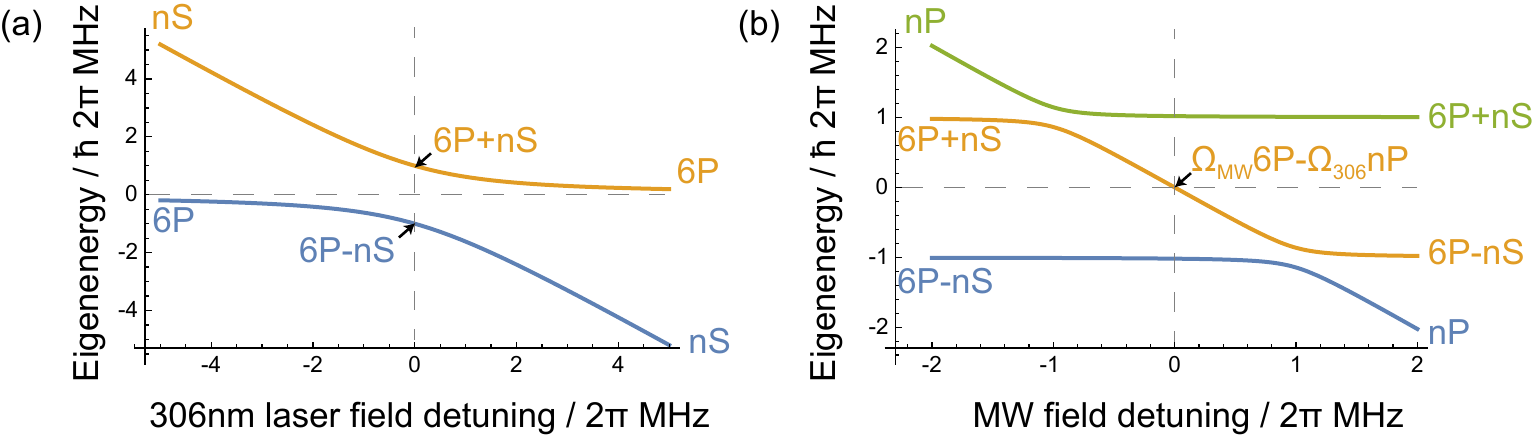}
\caption{Eigenenergies of the coupled atomic levels during probing of (a)~$nS_{1/2}$ states and (b)~$nP_{1/2}$ states using the coherent spectroscopy technique.
The weak 243\,nm probe field is resonant to the $4D_{5/2} \leftrightarrow 6P_{3/2}$ transition, and it optically pumps population from $4D_{5/2} \rightarrow 5S_{1/2}$ when the coupled system has an eigenstate with a $6P_{3/2}$ component at eigenenergy zero.
(a)~When the 306\,nm laser field is resonant to the $6P_{3/2} \leftrightarrow nS_{1/2}$ transition it causes an Autler-Townes splitting.
This prevents the weak 243\,nm probe field from optically pumping population from $4D_{5/2} \rightarrow 5S_{1/2}$.
Conversely when the 306\,nm laser field is detuned (by more than its coupling strength $\Omega_{306}$) the weak 243\,nm probe field optically pumps population from $4D_{5/2} \rightarrow 5S_{1/2}$.
(b)~When the MW field is not resonant to the $nS_{1/2} \leftrightarrow nP_{1/2}$ transition, the Autler-Townes splitting (due to the 306\,nm laser field) prevents the probe field from optically pumping population from $4D_{5/2} \rightarrow 5S_{1/2}$.
Conversely when the MW field is resonant an eigenstate [$N \left(\Omega_{\mathrm{MW}} \, 6P_{3/2} - \Omega_{306} \, nP_{1/2} \right)$, normalization constant $N$] with energy zero emerges and the weak, resonant 243\,nm laser field optically pumps population from $4D_{5/2} \rightarrow 5S_{1/2}$.
In the figure the coupling strength of $6P_{3/2}\leftrightarrow nS_{1/2}$ is $\Omega_{306}=2\pi\times 2\,\mathrm{MHz}$ and $nS_{1/2}\leftrightarrow nP_{1/2}$ is $\Omega_{\mathrm{MW}}=2\pi\times 0.4\,\mathrm{MHz}$.
The eigenstates written in the figure are not normalised, and the $J$ subscripts are omitted for clarity.
The eigenenergy spectrum (a) was measured and is shown in the Supplemental Material of \cite{Higgins2017b}.
}
\label{fig_supp_coh_spec}
\end{figure}

To study Rydberg $S$ states:
\begin{itemize}
\item{The ion is initialised in $4D_{5/2}$.}
\item{The weak 243\,nm laser field resonantly probes $4D_{5/2} \leftrightarrow 6P_{3/2}$.}
\item{The frequency of the stronger 306\,nm laser field is scanned.}
\item{If the 306\,nm laser field is detuned from resonance then ion population is quickly pumped by the 243\,nm probe laser field from $4D_{5/2}$ via $6P_{3/2}$ to the ground state $5S_{1/2}$.}
\item{If the 306\,nm laser field is near resonant to a $6P_{3/2} \leftrightarrow nS_{1/2}$ transition, then it introduces an Autler-Townes effect: $6P_{3/2}$ and $nS_{1/2}$ are coupled, and the new eigenstates are shifted. This means the 243\,nm laser field is no longer resonant with any transition, and the optical pumping rate from $4D_{5/2}$ via $6P_{3/2}$ to the ground state $5S_{1/2}$ is reduced. See Fig.~\ref{fig_supp_coh_spec}(a).}
\end{itemize}

The method relies on having a strong enough coupling induced by the 306\,nm laser field for an Autler-Townes effect to appear -- this requires a coupling strength $>\sim \Gamma_{6P} = 2\pi \times 4.5\,\mathrm{MHz}$.
This places a resolution limit of $\sim 2\pi \times 1\,\mathrm{MHz}$.
When searching for Rydberg $nS_{1/2}$ states the sensitivity is enhanced by using high 306\,nm intensities (to achieve large $\Omega_{306}$).

For spectroscopy of Rydberg $P_{1/2}$ states the resolution can be higher:
\begin{itemize}
\item{The ion is initialised in $4D_{5/2}$.}
\item{The weak 243\,nm laser field resonantly probes $4D_{5/2} \leftrightarrow 6P_{3/2}$.}
\item{The strong 306\,nm laser field resonantly couples $6P_{3/2} \leftrightarrow nS_{1/2}$ while the microwave field with variable detuning couples $nS_{1/2} \leftrightarrow nP_{1/2}$.}
\item{If the microwave field is detuned from resonance then ion population remains in $4D_{5/2}$, since the Autler-Townes splitting caused by the 306\,nm laser field effectively pushes the 243\,nm laser field off resonance, which means optical pumping from $4D_{5/2} \rightarrow 5S_{1/2}$ via $6P_{3/2}$ is prevented (the MW field only shifts the $nS_{1/2}$ level slightly and so the 306\,nm laser field is still near resonance).}
\item{If the microwave field is resonant to a $nS_{1/2} \leftrightarrow nP_{1/2}$ transition, then the coupled system has an eigenstate $N \left( \Omega_{\mathrm{MW}}\,6P_{3/2} - \Omega_{306}\,nP_{1/2} \right)$ which has zero energy ($N$ is a normalization constant), see Fig.~\ref{fig_supp_coh_spec}(b).
The weak 243\,nm probe field resonantly excites the $4D_{5/2} \leftrightarrow N \left( \Omega_{\mathrm{MW}}\,6P_{3/2} - \Omega_{306}\,nP_{1/2} \right)$ transition, and thus population is optically pumped from $4D_{5/2} \rightarrow 5S_{1/2}$.
}
\end{itemize}

The resonance linewidth $\Gamma_{\mathrm{spec}}$ depends on the couplings strengths $\Omega_{306}$ and $\Omega_{\mathrm{MW}}$
\begin{equation}
\Gamma_{\mathrm{spec}} \approx \frac{\Omega_{\mathrm{MW}}^2}{\Omega_{\mathrm{MW}}^2 + \Omega_{306}^2} \Gamma_{6P_{3/2}} + \frac{\Omega_{306}^2}{\Omega_{\mathrm{MW}}^2 + \Omega_{306}^2} \Gamma_{nP_{1/2}}
\end{equation}
A narrower linewidth can be obtained by using $\Omega_{\mathrm{MW}} \ll \Omega_{306}$ at the expense of excitation of more population to $nP_{1/2}$ and a larger risk of double ionization.
When searching for Rydberg $nP_{1/2}$ states the sensitivity is enhanced by using high MW intensities (to achieve large $\Omega_{\mathrm{MW}}$).


\section{Calculation of the response of Rydberg-P states to the quadrupole fields using diagonalization of the Hamiltonian} \label{sec_b}
The unperturbed Hamiltonian is $H_0$.
We have a magnetic field with strength 0.346\,mT, which introduces Zeeman shifts $H_Z$.
In our system in which the trap symmetry axis is collinear with the quantization axis (defined by the magnetic field) the quadrupole fields cause the perturbation
\begin{equation}
H_Q = \sum_i \sum_{j} e \left( \phi_{\mathrm{rf}} \cos{\Omega t} + \phi_{\mathrm{rad}} \right) |i\rangle \langle i | r^2 (Y_2^2 + Y_2^{-2}) | j \rangle \langle j | + e \phi_{\mathrm{ax}} |i\rangle \langle i | r^2 Y_2^0 | j \rangle \langle j |
\end{equation}
where the indices $i$ and $j$ describe the atomic states.

\begin{figure}[ht!]
\centering
\includegraphics[width=0.5\columnwidth]{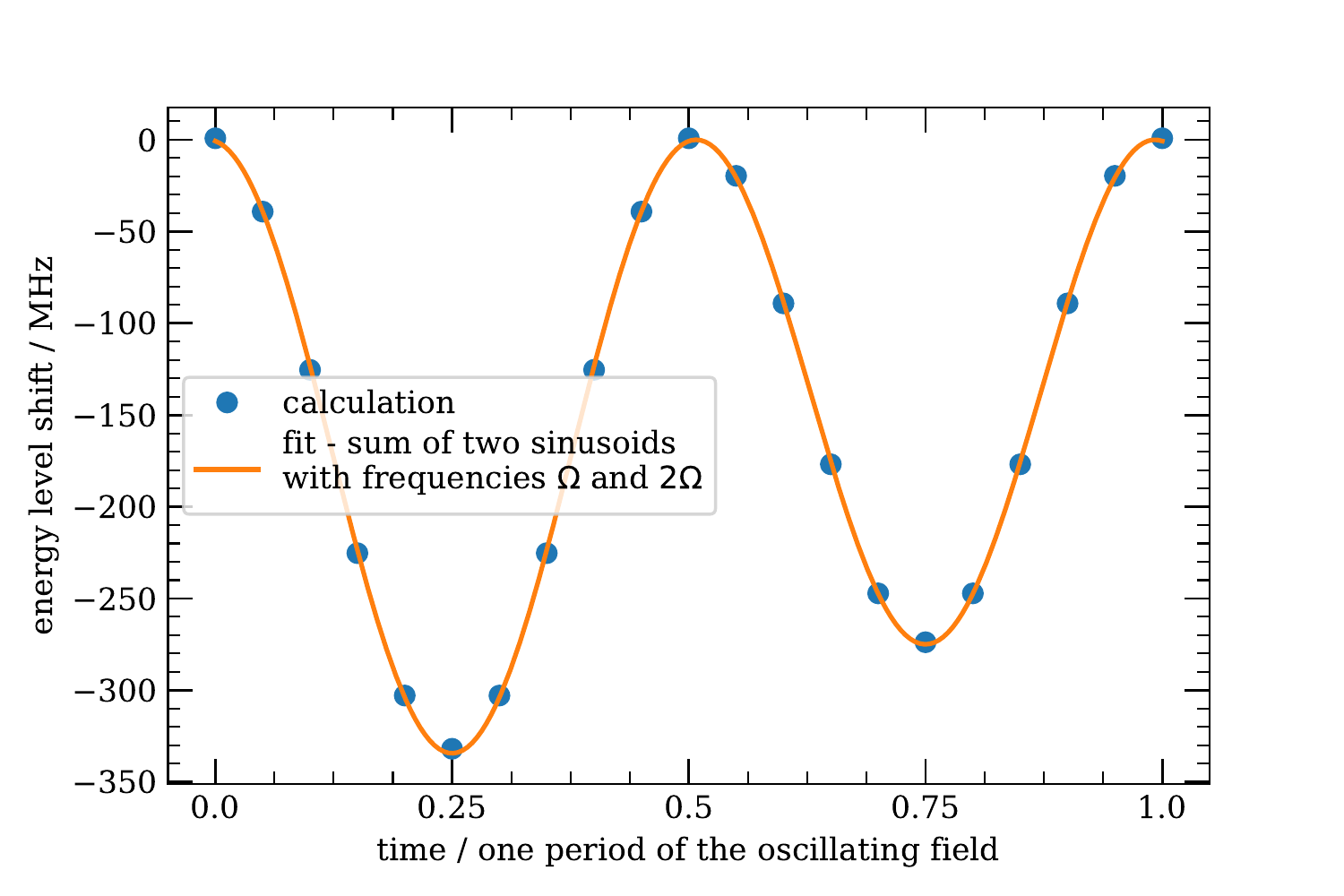}
\caption{Evolution of state $56P_{1/2}$ $m_J=\frac{1}{2}$ energy during one period of the oscillating quadrupole field. The oscillation has two components -- one with frequency $\Omega$ and the other with frequency $2\Omega$.
Trapping parameters: $\phi_{\mathrm{rf}}=10^9\,\mathrm{Vm^{-2}}$, $\phi_{\mathrm{rad}}=5\times10^7\,\mathrm{Vm^{-2}}$, $\phi_{\mathrm{ax}}=3\times10^7\,\mathrm{Vm^{-2}}$, $\Omega = 2\pi\times 18.11\,\mathrm{MHz}$.
}
\label{fig_supp_eigenenergy}
\end{figure}

We consider different instants in time across one period of the oscillating field.
At each instant we diagonalise the Hamiltonian.
We identify one of the eigenstates with the atomic state of interest -- $nP_{1/2} m_J=-\frac{1}{2}$.
We plot its eigenenergy at different times during one period in Fig.~\ref{fig_supp_eigenenergy}.
The average of the eigenenergy describes $E'_0$ (after accounting for the Zeeman shift), the $\Omega$-component of the oscillation in Fig.~\ref{fig_supp_eigenenergy} describes $E'_1$ and the $2\Omega$-component of the oscillation in Fig.~\ref{fig_supp_eigenenergy} describes $E'_2$.

\begin{figure}[ht!]
\centering
\includegraphics[width=0.5\columnwidth]{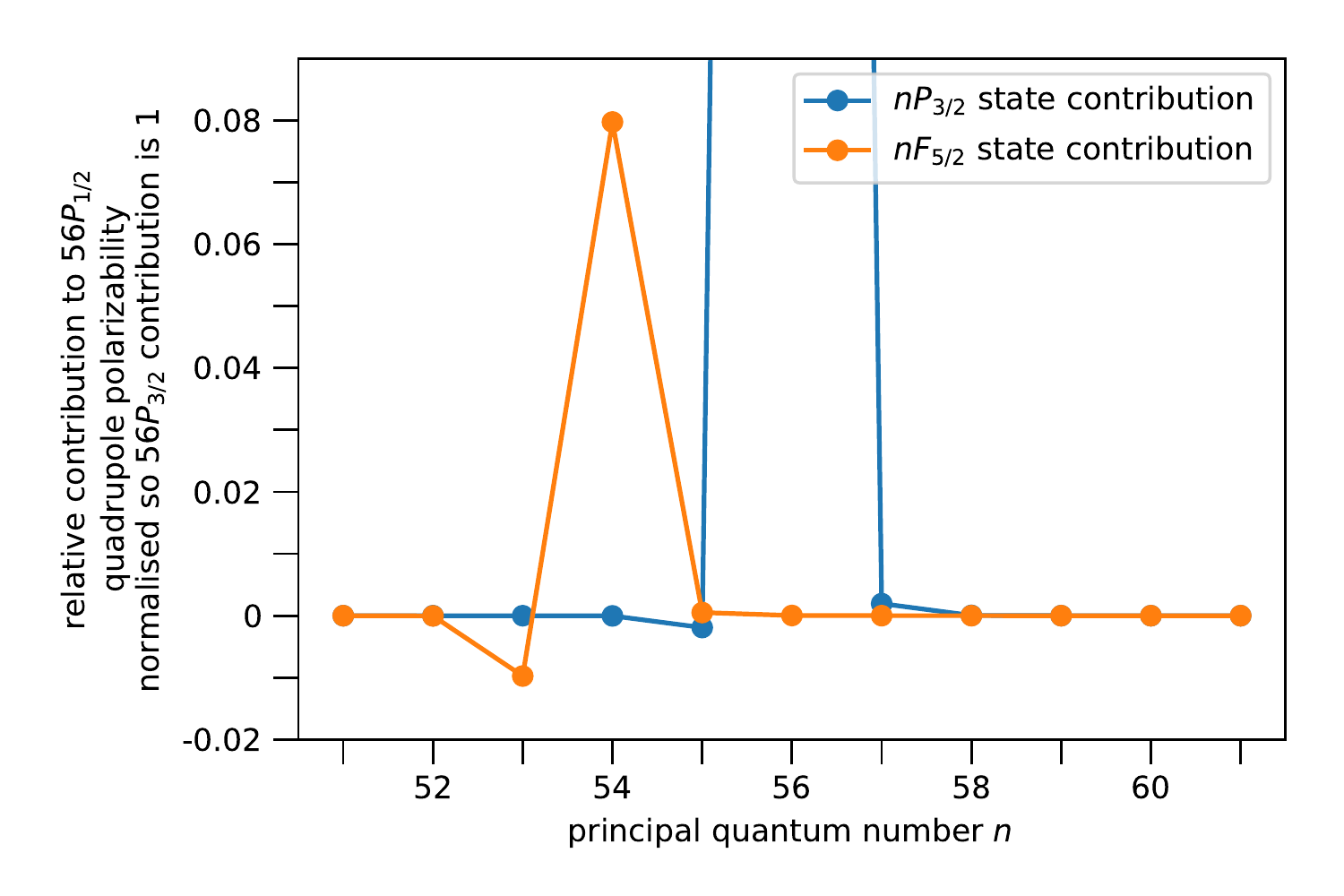}
\caption{Relative contributions to the quadrupole polarizability of $56P_{1/2}$.
The largest contribution is from $56P_{3/2}$.
The next largest contribution is from $54F_{5/2}$.
States with energies lower than $56P_{1/2}$ (such as $53F_{5/2}$) give negative contributions -- this is because of the energy difference in the denominator of the second-order perturbation calculation in Eq.~\ref{eq_second_order_pert}.
}
\label{fig_supp_contributions}
\end{figure}

In this calculation our basis consists of the 20 states:
\begin{itemize}
\item{The two sublevels of $nP_{1/2}$.}
\item{The four sublevels of $nP_{3/2}$.}
\item{The six sublevels of $(n-2)F_{5/2}$.}
\item{The eight sublevels of $(n-2)F_{7/2}$.}
\end{itemize}
This is because the quadrupole fields cause coupling between states with $\Delta L = 0$ or $\Delta L =\pm 2$, and the $(n-2)F$ states are the closest $F$ states to $nP$, and contribute the most to the quadrupole coupling.
This is justified in Fig.~\ref{fig_supp_contributions} where the relative contributions to the quadrupole polarizability of $56P_{1/2}$ are shown; this was found using the perturbative approach of Section~\ref{sec_a}.

\begin{figure}[ht!]
\centering
\includegraphics[width=0.5\columnwidth]{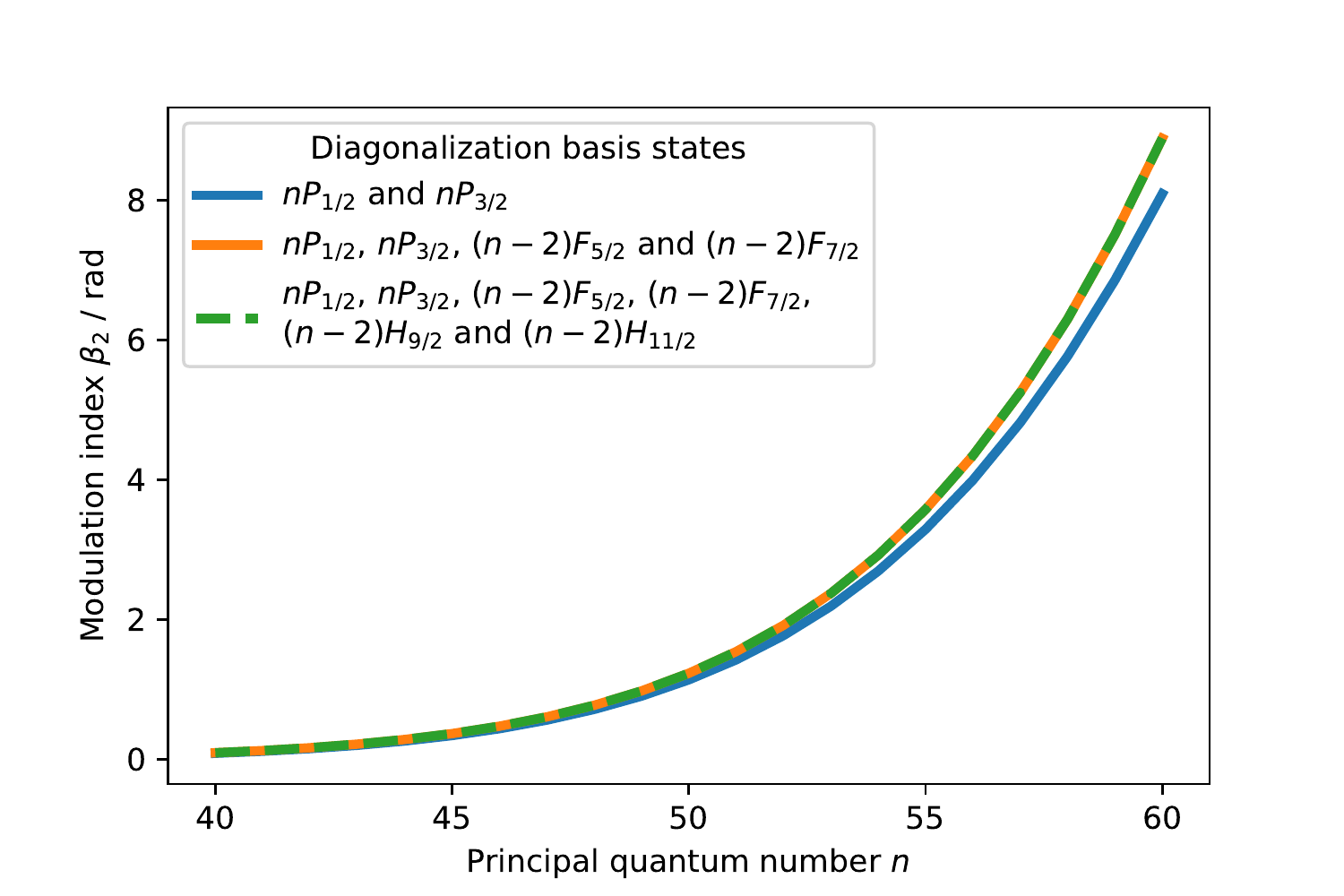}
\caption{Effect of changing basis states used in diagonalization calculation. The calculated modulation index $\beta_2$ is shown for different Rydberg $P_{1/2}$ states, when the trap settings from Fig.~2 of the main text are used.
Inclusion of the nearest $H$ states causes a negligible change in the calculation results. Restriction of the basis to only $P$ states leads to inaccurate predictions.
}
\label{fig_supp_H_states}
\end{figure}

Quadrupole-mixing between high-$L$ states could conceivably produce effects on $nP_{1/2}$ for the field strengths considered in this work.
To investigate this we expanded the basis to include the $(n-2)H_{9/2}$ and $(n-2)H_{11/2}$ states.
This had a negligible impact on the calculation results (see Fig.~\ref{fig_supp_H_states}), this means we do not need to consider effects of high $L$ states on $nP_{1/2}$.

We also tried restricting the basis to only $nP_{1/2}$ and $nP_{3/2}$ states, but this caused significant changes in the predictions as shown in Fig.~\ref{fig_supp_H_states}.
This is unsurprising, given the contribution of $54F_{5/2}$ states to the $56P_{1/2}$ quadrupole polarizability in Fig.~\ref{fig_supp_contributions}.
And thus we did not use this restricted basis for the calculations used in the main text.

\section{Modulation of the Rydberg state's energy causes sidebands in Rydberg-excitation spectra which are described by Bessel functions} \label{sec_modulation}
This description follows the approach of \cite{Feldker2015}.
The quadrupole fields shift the Rydberg energy levels to
\begin{equation}
E' = E'_0 + E'_1 \cos{\Omega t} + E'_2 \cos{2 \Omega t}
\end{equation}
For a Rydberg-excitation transition, the resonance frequency varies in time according to
\begin{equation}
\omega(t) = \omega'_0 + \frac{E'_1}{\hbar} \cos{\Omega t} + \frac{E'_2}{\hbar} \cos{2 \Omega t}
\end{equation}
The field resonant to the transition is
\begin{equation}
\mathcal{E}(t) \sim e^{-i \int \omega(t') dt'} = \mathrm{exp}\left(- i \omega'_0 t - i \frac{E'_1}{\hbar \Omega} \sin{\Omega t} - i \frac{E'_2}{2 \hbar \Omega} \sin{2 \Omega t}\right)
\end{equation}
Using the Jacobi-Anger expansion
\begin{equation}
e^{i \beta \sin{\theta}} = \sum_n J_n(\beta) e^{i n \theta}
\end{equation}
then
\begin{equation}
\mathcal{E}(t) \sim e^{-i \omega'_0 t} \sum_m J_m\left(\frac{E'_1}{\hbar \Omega}\right) e^{-i m \Omega t} \sum_n J_n\left(\frac{E'_2}{2 \hbar \Omega}\right) e^{-i 2 n \Omega t}
\end{equation}
and thus peaks $\omega'_0 + m \Omega + 2 n \Omega$ result, for integers $m$ and $n$.

\section{Sidebands in Rydberg-P excitation spectra depend on quadrupole response of both Rydberg-S and Rydberg-P states}
We probe $nP_{1/2}$ states using the coherent spectroscopy technique described in Section~\ref{sec_coherent_spectroscopy}.
We use a 243\,nm laser field to resonantly couple $4D_{5/2}$ and $6P_{3/2}$, and a 306\,nm laser field to resonantly couple $6P_{3/2}$ and $nS_{1/2}$.
The $nS_{1/2}$ energy level oscillates due to the oscillating quadrupole field of the trap, however the 306\,nm laser field is at a fixed frequency, resonant to the carrier transition.
Within a Floquet picture, the 306\,nm laser field couples $6P_{3/2}$ to the $nS_{1/2}$-Floquet-state that has zero quasi-energy.
A microwave field couples $nS_{1/2}$ and $nP_{1/2}$; the energies of both levels oscillate in phase due to the oscillating quadrupole field of the trap.
The energy difference between the transition energy and a microwave photon oscillates at $2\Omega$, with amplitude $E_{\mathrm{nP}}-E_{\mathrm{nS}}$, where $E_x$ is the oscillation amplitude of state $x$.
As a result, the modulation index describing the spectral sidebands is $(E_{\mathrm{nP}}-E_{\mathrm{nS}})/(2 \hbar \Omega)$ (the way a modulated energy level gives rise to spectral sidebands is explained in Section~\ref{sec_modulation}).

\begin{figure}[ht!]
\centering
\includegraphics[width=0.7\columnwidth]{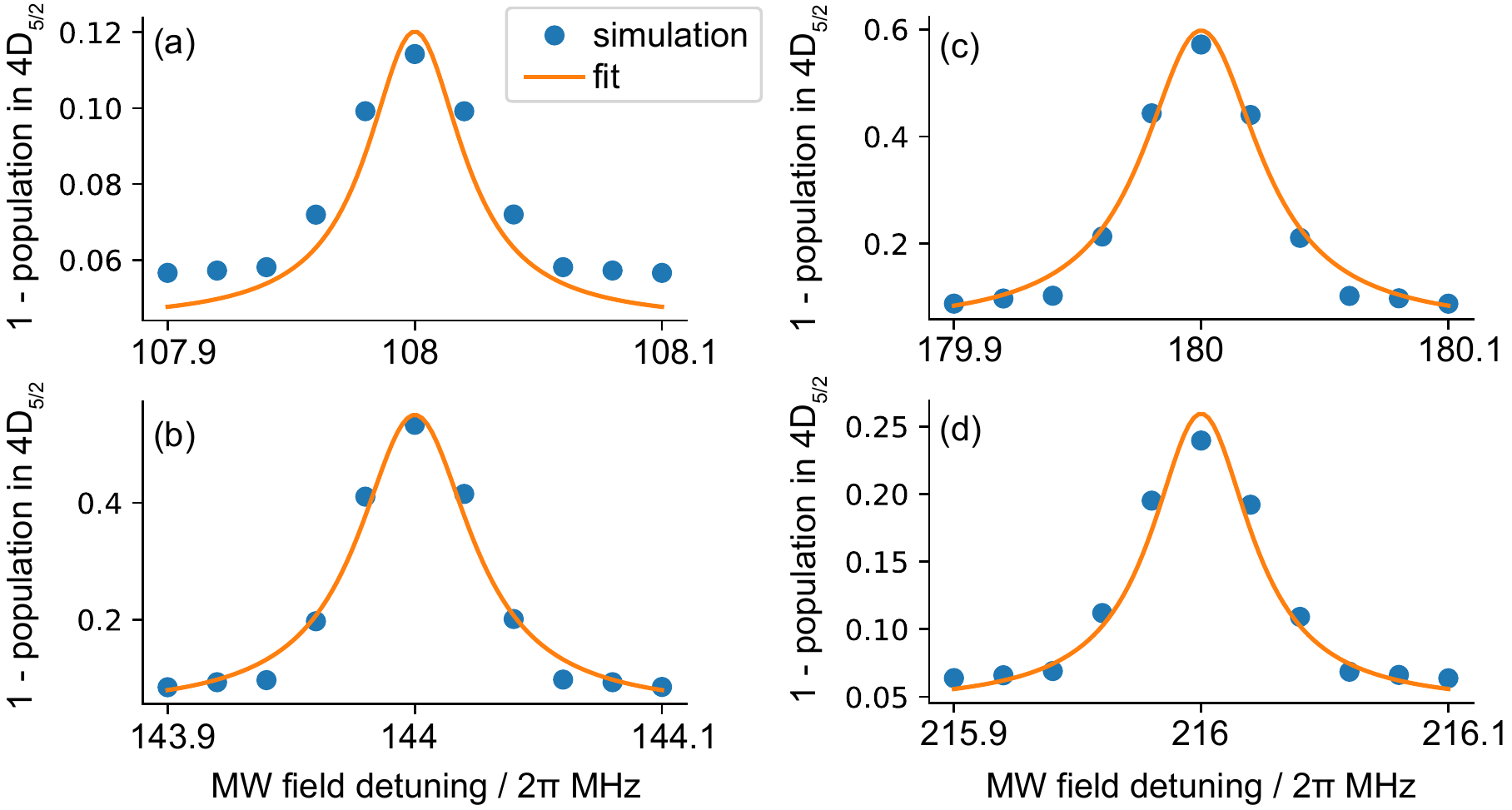}
\caption{Simulated excitation spectrum for an $nP_{1/2}$ state using the coherent spectroscopy technique of Section~\ref{sec_coherent_spectroscopy}.
The $nP_{1/2}$ is modulated at frequency $2\Omega=2\pi\times$36\,MHz with amplitude $E_{nP}=\hbar\times2\pi\times$108\,MHz, while the $nS_{1/2}$ state is modulated in phase with amplitude $E_{nS}=\hbar\times2\pi\times$10.8\,MHz.
Sidebands appear at multiples of $2\Omega$.
The sideband amplitudes are described by the modulation index $\beta_2=\frac{E_{nP}-E_{nS}}{2\Omega}=2.7\,\mathrm{rad}$.
(a) Carrier resonance, (b) first sideband, (c) second sideband and (d) third sideband.
The carrier resonance is detuned by $2\pi\times$108\,MHz due to the oscillating quadrupole field according to Eq.~4 of the main text ($E_0'=E_2'$).
}
\label{fig_supp_modulation_1}
\end{figure}

We check this by simulating the five-level system (\{$4D_{5/2}$, $6P_{3/2}$, $nS_{1/2}$, $nP_{1/2}$ and $5S_{1/2}$\}) with a resonant $4D_{5/2} \leftrightarrow 6P_{3/2}$ coupling, a resonant $6P_{3/2} \leftrightarrow nS_{1/2}$ coupling, and a field coupling $nS_{1/2} \leftrightarrow nP_{1/2}$.
The energies of the levels $nS_{1/2}$ and $nP_{1/2}$ are modulated in phase with frequency $2 \Omega$.
The detuning of the microwave field is scanned, and the excitation spectrum is shown in Fig.~\ref{fig_supp_modulation_1}.

\begin{figure}[ht!]
\centering
\includegraphics[width=0.5\columnwidth]{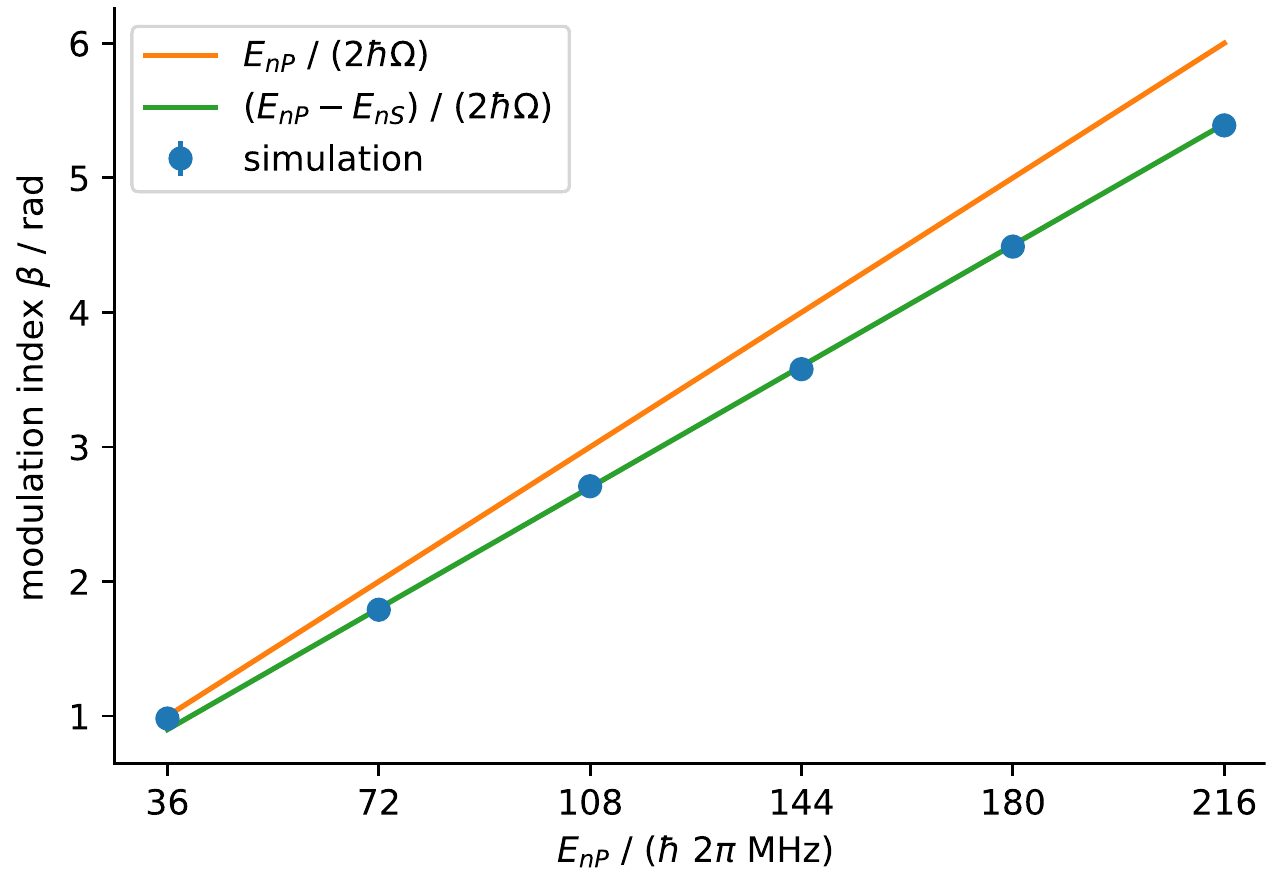}
\caption{Spectra as in Fig.~\ref{fig_supp_modulation_1} are simulated as the modulation amplitudes $E_{nP}$ and $E_{nS}$ are varied ($E_{nS} = \frac{1}{10} E_{nP}$.
The modulation index $\beta$ describing the spectra satisfies $\beta=\frac{E_{nP}-E_{nS}}{2\Omega}$.
}
\label{fig_supp_modulation_2}
\end{figure}

The amplitudes with which the energy levels are modulated is varied, and the modulation index is extracted from each excitation spectrum.
The dependence of the modulation index on the amplitudes of energy level modulation is show in Fig.~\ref{fig_supp_modulation_2}.
The modulation index depends on the modulation of both the $nS_{1/2}$ and $nP_{1/2}$ levels.

\section{Interplay of two modulations causes an asymmetric spectrum}
\begin{figure}[ht!]
\centering
\includegraphics[width=\columnwidth]{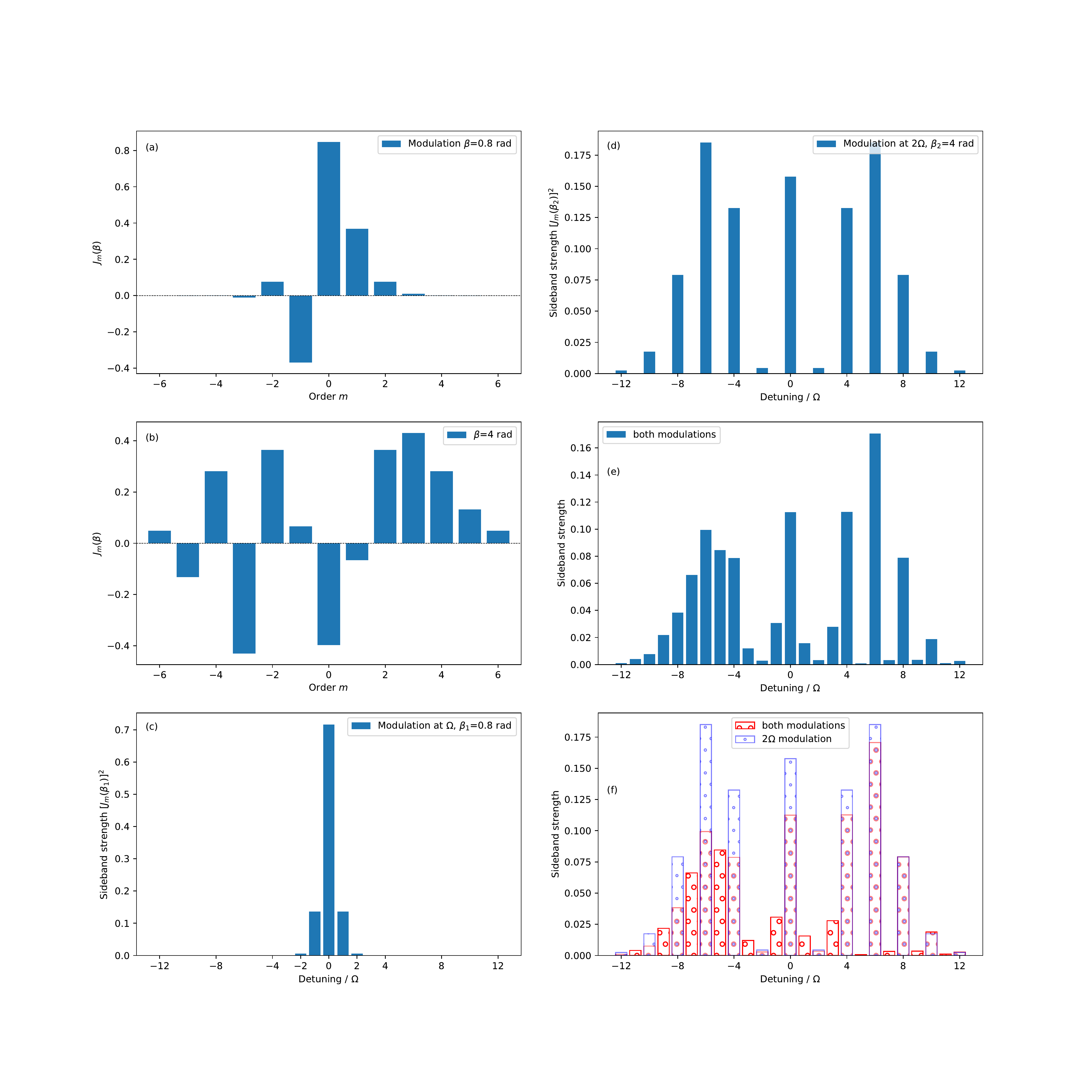}
\caption{The modulation with frequency $\Omega$ and modulation index $\beta_1$ and the modulation with frequency $2\Omega$ and modulation index $\beta_2$ together give rise to an asymmetric spectrum.
}
\label{fig_supp_asymmetry}
\end{figure}

There is an interplay between the modulation at $\Omega$ with modulation index $\beta_1$ and the modulation at $2\Omega$ with modulation index $\beta_2$.
A similar interplay of modulations was observed in \cite{Feldker2015}.
If there was only $\Omega$ modulation, then the $p^{\mathrm{th}}$ sideband at detuning $p\times\Omega$ would have strength strength $\sim [J_p(\beta_1)]^2$ ($J_p(\beta_1)$ is the $p^{\mathrm{th}}$-order Bessel function of the first-kind, with modulation index $\beta_1$), see Fig.~\ref{fig_supp_asymmetry}(c).
If there was only $2\Omega$ modulation, then the $q^{\mathrm{th}}$ sideband at detuning $q\times2\Omega$ would have strength strength $\sim [J_q(\beta_2)]^2$, see Fig.~\ref{fig_supp_asymmetry}(d).

The interplay between the modulations causes a coupling at $r \times \Omega$ with strength
\begin{equation}
\sum_p \sum_{q} J_p(\beta_1) J_q(\beta_2) \delta_{p+2q,r}
\end{equation}
where $\delta_{i,j}$ is the Kronecker delta.
This leads to a sideband at $r \times \Omega$ with strength
\begin{equation}
\left(\sum_p \sum_{q} J_p(\beta_1) J_q(\beta_2) \delta_{p+2q,r}\right)^2
\end{equation}
As shown in Fig.~\ref{fig_supp_asymmetry}(e) and (f) this results in an asymmetric spectrum. The asymmetry is due to the following property of Bessel functions
\begin{equation}
J_{-m}(\beta) = (-1)^m J_m(\beta)
\end{equation}


\end{document}